# Negative Index Makes a Perfect Time-Domain Lens, Generating Slow Playback of Ultrafast Events

Oded Schiller[1,3 †], Yonatan Plotnik[3 †], Guy Bartal[1] and Mordechai Segev[1,2,3]

1. *Department of Electrical and Computer Engineering, Technion, Haifa 32000, Israel*
2. *Physics Department, Technion, Haifa 32000, Israel*
3. *Solid State Institute, Technion, Haifa 32000, Israel*

*† Contributed Equally*

**Abstract**

We explore the effects of incorporating negative index materials into the physics of time-varying media and find that changing the refractive index from positive to negative creates a perfect time-reversed wave: a perfect time-domain lens. Unlike other mechanisms of phase conjugation, the perfect time-domain lens time-reverses both the propagating waves and the evanescent part of the spectrum. Moreover, we find that the time-reversed wave can be slowed down or accelerated, depending on the refractive index ratio. We show that this effect remains strong even when the refractive index varies arbitrarily slow, in sharp contradistinction to time-reflection – which necessitates large index changes at sub-cycle rates. This is the first avenue found to yield significant negative-frequency waves using a temporal interface without the need for sub-cycle modulation or impedance matching. The effect can be used to record extreme ultrafast information and subsequently play it backwards at a slow rate, and vice-versa.

Wave propagation in time-varying media is currently attracting considerable attention, following recent experimental progress on dramatically changing the refractive index within a single cycle, in microwaves [1,2] and at optical frequencies [3]. Its most fundamental aspects are time-refraction and time-reflection [4–6]. To understand these, consider a time-varying medium where the refractive index changes abruptly from $n_1$ to $n_2$, creating a time-interface. When an electromagnetic (EM) wave of frequency $\omega_1$ experiences a time-interface in a homogeneous medium, the momentum of the wave– its wavevector $\vec{k}$, is conserved. Consequently, the wave splits into a time-refracted wave and a time-reflected wave. The time-refracted wave moves in the same direction as the original wave but experiences a frequency shift from $\omega_1$ to $\omega_2$, because $k = \omega_1 n_1 / c = \omega_2 n_2 / c$, $c$ being the vacuum speed of light. The time-reflected wave is also frequency shifted, but to the negative frequency of $-|\omega_2|$. Since it is impossible to evolve backwards in time, this time-reflected wave travels in the opposite direction in space with a conjugate amplitude. The time-reflected wave is closely related to phase-conjugation [7]. Based on time-reflected and time-refracted waves, a plethora of exciting phenomena can be explored, ranging from breaking reciprocity [8], inverse prisms [9] and temporal aiming [10] to extreme non-Hermitian energy transfer [11] and time-domain bound states in the continuum [12]. One of the most interesting phenomena in time-varying media are Photonic Time-Crystals (PTCs) [6,13,14]. A PTC is a medium with a refractive index that changes periodically in time, giving rise to multiple time-refractions and time-reflections, which interfere to yield a band structure consisting of momentum bands separated by gaps. In these momentum gaps the frequency is complex, leading to modes that grow (or decay) exponentially in time, drawing energy from the index modulation (or transferring energy to the modulator source). PTCs have many intriguing features, among them non-trivial topology [15], new opportunities for nonlinear optics [16–19] and breaking the Rozanov bound for absorption [20]. Moreover, because the gain in the gap modes of a PTC does not rely on atomic resonances, it can be used to create lasers [21] at frequencies where lasers are hard to make. PTCs also exhibit interesting quantum effects [21–24], and reach further – to temporally-disordered systems [25,26], and time-quasicrystals [27–29]. PTCs were recently realized experimentally in microwaves, creating a non-resonant source of broadband gain [30].

Experimentally, while time-refraction was observed in optical systems [3,31,32], time-reflection was thus far demonstrated in water waves [33], ultracold atoms [34], microwaves [1,2] and synthetic dimension systems [35,36]. Despite recent experimental advancements displaying large refractive index changes in the optical single-cycle regime [3], time-reflection at optical frequencies has never been observed. There are two main reasons for that. First, to create a measurable time-reflected wave, the index change must be of order of unity. Second, the index change should occur within a sub-cycle of the optical wave, i.e., a few femtoseconds. These conditions are extremely challenging because all known mechanisms to change the refractive index are either too weak (e.g., the optical Kerr effect) or too slow (such as acousto-optic or nonlinearities in liquid crystals). A slow response time results in a vanishingly weak time-reflected wave, with amplitude decaying exponentially with the ratio between the transition time and the period of the EM wave. In fact, response time of several cycles already reduces the time-reflected amplitude by orders of magnitude. Hence, it is natural to ask whether it is possible to generate a significant negative-frequency wave from a time-interface, without the need for sub-cycle modulation, and if so – find a suitable mechanism.

Here, we show that a time-interface, where the refractive index varies from positive to negative, gives rise to a strong negative-frequency (time-reversed) wave, even when the index change occurs arbitrarily slowly. This is the first mechanism producing a strong negative-frequency wave induced by a time-interface without the need for sub-cycle modulation. We find that, during an index change from $n_1 > 0$ to $n_2 < 0$, the time-reflection and time-refraction switch roles: the negative frequency wave is created by time-refraction and not by time-reflection. This effect can be used to create a perfect time-domain lens, which enables slowed-down time-reversed temporal evolution of ultrafast information: one can capture ultrafast phenomena that are not accessible using conventional measurement devices. This idea paves the way for a range of applications relying on the ability to capture ultrafast events with a system that has a long capture time.

Our proposed system incorporates negative index material (NIM) into time-varying media. NIM (or left-handed martials) have both $\varepsilon$ and $\mu$ negative, endowing the waves propagating within them with unique properties [37–41]. Perhaps the most important application of NIM is for perfect imaging [37], including the evanescent waves, which breaks the diffraction limit [38]. In fact, NIM inspired further ideas such as super-lens [42], hyper-lens [43,44], and invisibility cloaks [45–48]. However, using NIM in the context of temporal boundaries has been limited to instantaneous changes of the refractive index [49–51] and did not consider slow changes and their ability to create conjugate evanescent waves and sub-wavelength imaging.

Consider a spatially homogenous material where the permeability and permittivity ($\varepsilon(t), \mu(t)$) change in time. For simplicity, assume both $\mu(t)$ and $\varepsilon(t)$ are either positive or negative but not mixed cases, such that $n^2(t) = \mu(t)\varepsilon(t)$ is always real (for mixed cases see [52]). Maxwell's equations yield

$$\nabla \times \vec{B} = \mu_0 \mu(t) \partial_t \vec{D}, \qquad \nabla \times \vec{D} = -\varepsilon_0 \varepsilon(t) \partial_t \vec{B} \qquad (1)$$

Because of spatial homogeneity, the wavevector $\vec{k}$ is conserved, and any general solution can be expressed as a superposition of plane waves.

First, we explore the case where a wave experiences an abrupt time-interface, and the medium changes from $\varepsilon_1, \mu_1$ to $\varepsilon_2, \mu_2$, where either one or both of those materials can be a NIM. The displacement field, $\vec{D}$, before the time boundary, is of the form:

$$\vec{D}(t < 0, z) = \hat{x} D_0 e^{-ikz} e^{i\omega_1 t} \tag{2}$$

Where the dispersion relation yields $\omega_1 = \frac{ck}{n_1}$ with $n_1 = sign(\mu_1)\sqrt{\varepsilon_1 \mu_1}$. After the time-interface, due to conservation of $\vec{k}$ and the new dispersion relation, $\vec{D}$ must take the form:

$$\vec{D}(t > 0, z) = \hat{x} D_0 e^{-ikz} (\tau e^{i\omega_2 t} + \Gamma e^{-i\omega_2 t}) \tag{3}$$

Where $\omega_2 = \frac{ck}{n_2}$, $n_2 = sign(\mu_2)\sqrt{\varepsilon_2 \mu_2}$, $\tau, \Gamma$ being the time-refraction and time-reflection coefficients, respectively. From this we find:

$$\omega_2 = \frac{n_1}{n_2} \omega_1 \tag{4}$$

This relation defines the frequency of both refracted and reflected wave, but the frequency of the reflected wave gets a negative sign. When both $n_1$ and $n_2$ are positive, $\omega_2$ and $\omega_1$ have the same sign, thus the time-reflected wave appears as a wave of negative frequency propagating backwards in space with a conjugate phase (Fig. 1a). When $n_1$ and $n_2$ have opposites signs, $\omega_2$ has an opposite sign to $\omega_1$. This means that, when the time boundary is between positive and negative index materials, the time-reflection and time-refraction exchange roles (Fig. 1b). Now, the negative frequency wave is the transmitted wave – which is now propagating backwards in space with a conjugate phase, and the positive frequency wave is the time-reflected wave, which is propagating forwards in space. This is the key insight for the rest of the derivation.

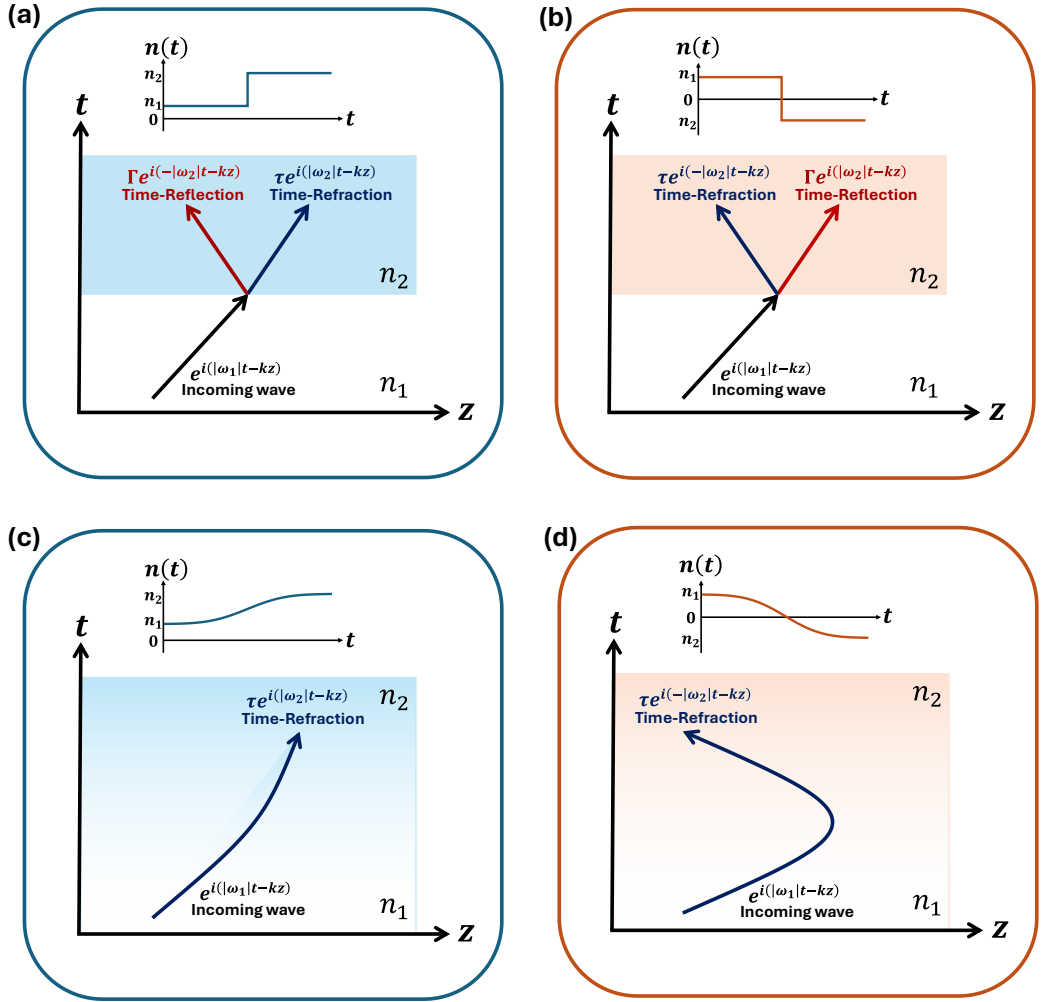

*Fig. 1.* **Time boundaries with positive and negative refractive indices.** (a) Abrupt time-boundary where both $n_1, n_2 > 0$. The time-refracted wave (blue) has positive frequency and continues in the same direction as the incoming wave. The time-reflected wave (red) has negative frequency and travels in the opposite direction. (b) Abrupt time-boundary between $n_1 > 0$ and $n_2 < 0$. The time-refraction and the time-reflection exchange roles. Now, the time-refracted wave (blue) has negative frequency and travel in the direction opposite to the incoming wave, while the time-reflected wave (red) has positive frequency and travels in the same direction as the incoming wave. (c) Slow time-boundary where both $n_1, n_2 > 0$. Because the time boundary is slow, only the time-refracted positive frequency wave is created. (d) Slow time-boundary where $n_1 > 0$ and $n_2 < 0$. Because the time boundary is slow, only the time-refracted wave is created, and because the time-refracted and time-reflected waves exchange roles, now only the negative frequency wave is created.

Deriving the transmission and reflection coefficients, using conservation of wavevector and continuity of $D, B$ at temporal interfaces, yields:

$$\tau = \frac{1}{2}\left(1 + \frac{Z_1}{Z_2}\right), \qquad \Gamma = \frac{1}{2}\left(1 - \frac{Z_1}{Z_2}\right) \tag{6}$$

Where, $Z_\alpha = \sqrt{\frac{\mu_\alpha}{\varepsilon_\alpha}}$ is the relative wave impedance of the respective medium. These coefficients do not depend on the sign of the indices. By examining these coefficients, we notice another counterintuitive effect: at a time-boundary switching from a positive index to a negative index, the negative frequency wave always has a larger amplitude than the positive frequency wave. This is also the opposite of what happens when the $n_1$ and $n_2$ have the same sign, where the time-refraction always has a larger amplitude than the time-reflection.

The most extreme case is the impedance-matched case, $Z_1 = Z_2$ switched from $n_1 > 0$ to $n_2 < 0$, Fig. 2. In this case, the amplitude of the time-reflected wave is completely nulled, leaving only the time-refracted negative-frequency wave, which evolves backwards in space with a conjugated phase. Thus, irrespective of the polarization of the wave (given the medium is isotropic) and for every wavevector that experiences this change in $\varepsilon$ and $\mu$, the temporal evolution is perfectly reversed - because of the minus sigh in the temporal phase. Furthermore, since the problem is linear in the wave amplitudes, where any initial field can be decomposed into a superposition of plane waves with different wavevectors and polarizations, the time-boundary time-reverses any arbitrary choice of initial field distribution. This makes a perfect time-domain lens: a device that perfectly reverses the temporal evolution of the EM fields, as demonstrated by the simulation in Fig 2.

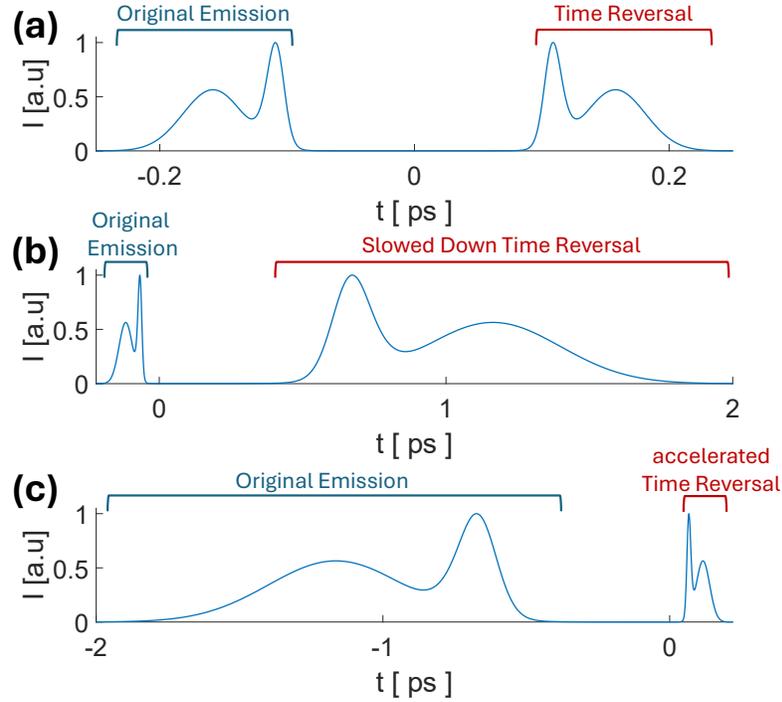

*Fig. 2.* **FDTD simulation of EM emission generated by a point source placed inside of a perfect time-domain lens and emitting a nontrivial temporal pulse.** (a) The source is placed in a perfect time-domain lens where $\left|\frac{n_1}{n_2}\right| = 1$. An exact time reversed copy is generated by the time-interface, evolving at the same rate as the incident field. Both wavepackets have temporal duration of 100[fs]. (b) The source is placed in a perfect time-domain lens where $\left|\frac{n_1}{n_2}\right| = \frac{1}{10}$, hence the time reversal is slowed down. The original pulse has duration of 100[fs], while the time-reversed pulse is slowed down to 1[ps] duration. (b) The source is placed in a perfect time-domain lens where $\left|\frac{n_1}{n_2}\right| = 10$, thus the time reversal is accelerated. The original pulse has duration of 1[ps], while the time reversed version is compressed to 100[fs].

Intriguingly, depending on the ratio $\left|\frac{n_1}{n_2}\right|$, which determines the magnitude of the frequency shift, the temporal evolution can be slowed down or accelerated. When $\left|\frac{n_1}{n_2}\right| < 1$ the evolution is slowed down, because the negative frequency is red-shifted compared to the original frequency. On the other hand, if $\left|\frac{n_1}{n_2}\right| > 1$ the evolution is accelerated, because the negative frequency is blue-shifted. This introduces an opportunity: using a time-boundary between an ordinary material and a NIM, an ultra-fast phenomenon can be played backwards at a much slower rate. This enables measuring ultrafast phenomena with conventional slow measuring devices. The opposite creates a blue-shifted accelerated version, allowing pulse compression. Figure 2 presents the temporal shape of the displacement field $\vec{D}$, where a point source emits a nontrivial waveform. The emission of the point source is perfectly time-reversed in both accelerated and slowed down version. In fact, we find that the time-reversed emission does not distinguish between propagating waves and evanescent waves: varying the refractive index from positive to negative time-reverses all waves exiting in the medium, propagating as well as evanescent. Proof of that is given in [58]. An example of this is shown in Fig. 3(a-c), where the simulated spatial resolution is arbitrarily high.

Next, we study the more physical case where the change from $n_1$ to $n_2$ is gradual and has a non-zero transition time, $T_{tr}$. Recalling results from time-modulation with only positive indices [4–6], we note that time-reflection requires (1) modulation faster than a single cycle of the EM wave and (2) good impedance mismatch. Failing to achieve these extremely challenging two conditions results in weak to vanishing time reflection. However, as shown above, when $n_1$ and $n_2$ have opposite signs, the time-refraction and time-reflection exchange roles: the negative frequency wave is now the transmitted wave, and the positive frequency wave is the reflected wave. This becomes especially important when the index change occurs gradually, because now the negative frequency (now time-refracted) wave - which is now propagating backwards in space - dominates. The amplitude of the transmitted wave is always high, even when the change in the refractive index occurs adiabatically and arbitrarily slowly. If the change is abrupt, both the time-refracted and the time-reflected waves exist, and the magnitude of the perfect time-domain lens is based on impedance matching. However, when change in $\varepsilon(t)$ and $\mu(t)$ is slow, all the power goes to the time-refracted wave, even without impedance matching ($Z_1 \neq Z_2$), Fig (3). This is important because achieving perfect impedance matching is challenging. This means that, using a gradual temporal change where $n(t)$ changes signs, will results in a strong time-refracted of negative-frequency even without impedance matching ($Z_1 \neq Z_2$). This is *a novel mechanism for creating the negative frequency wave without the need for sub-cycle switching of the refractive index*.

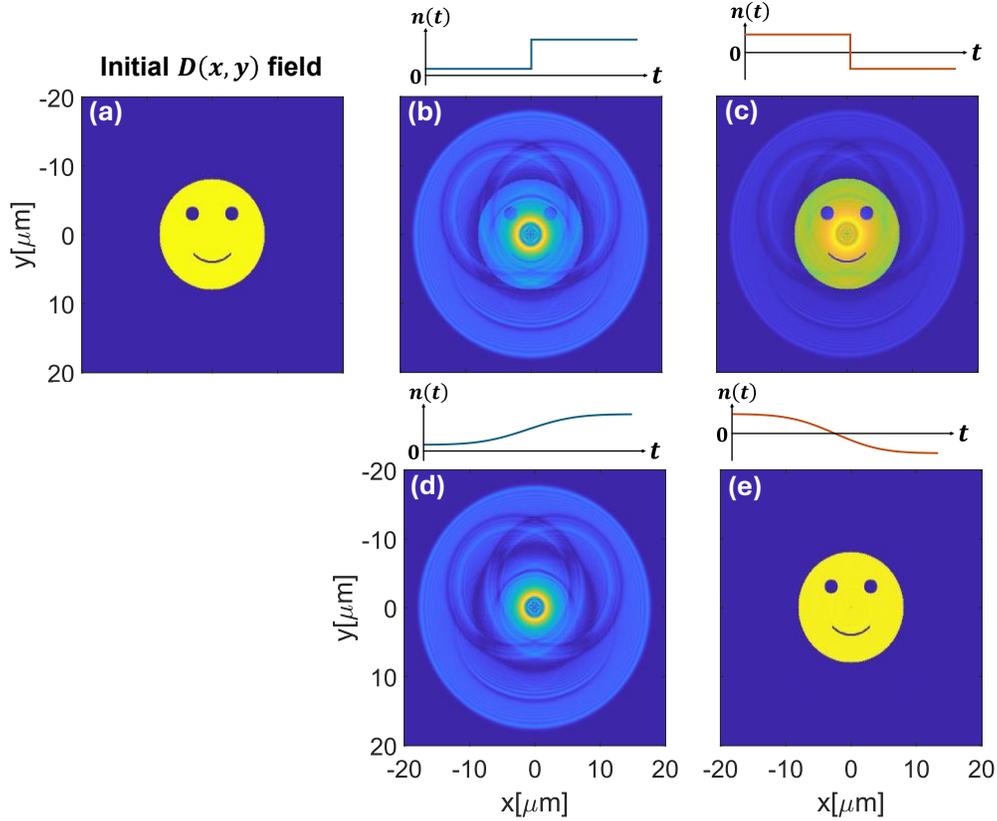

*Fig. 3.* **FDTD simulation of time-boundaries of different kinds and their effect on the generation of the negative frequency wave.** Here, all the time boundaries are impedance mismatched. (a) Initial amplitude of the field distribution. (b) Field after an abrupt time boundary with $n_1, n_2 > 0$; the amplitude of the negative frequency wave is much lower than that of the positive frequency wave. (c) Feld distribution after an abrupt time boundary where $n_1 > 0$ and $n_2 < 0$; the amplitude of the negative frequency wave is much bigger than that of the positive frequency wave. (d) Field distribution after a slow time boundary with $n_1, n_2 > 0$; the amplitude of the negative frequency wave is exponentially smaller than that of the positive frequency wave. (e) Field distribution after a slow time boundary with $n_1 > 0$ and $n_2 < 0$; the amplitude of the negative frequency wave dominates while the amplitude of the positive frequency wave is exponentially smaller.

Figure 3 shows that, when transition times is three periods and longer, the intensity of the negative-frequency wave is dominant. This raises a natural question: what happens to the temporal resolution of a pulse, when the medium undergoes a slow transition? Evidently, as shown in Figs. 2,3, the time-variation creates a phase-conjugate replica of the pulse, even though the switching is

Page 10 of 17

much slower than the pulse duration. This seems to be counterintuitive: how can a slow-responding device capture information that varies faster than its response time? Surely, a video camera can record only events evolving slower than its response time, not faster, so how come the slow switching from $n_1>0$ to $n_2<0$ can capture events much faster than the switching time, and broadcast them back with a conjugated amplitude? This mystery is resolved by understanding what is happening to the waves when the refractive index is varied. Physically, the ultrafast dynamics is "stored" by the presence of the propagation-evolution in the medium, hence – when the index is varied - even arbitrarily slowly – the index change causes the medium to "broadcast" all the spatial information present in the medium (both propagating and evanescent waves) and the temporal information, including features varying much faster than the variation time of the index. When the entire material undergoes a change in the refractive index, the time-refracted and time-reflected wave are generated, and the entire medium acts as "distributed antennae", broadcasting both the forward and backward propagating waves. This marks a notable difference from a camera or a mirror, including a phase-conjugate mirror, which can record only the information at a single plane; here, the information is "recoded" in the entire medium in which it propagates. In essence, the temporal events are not "captured" by the index variation, but by the presence of the wave in the medium. To prove our insight, it is instructive to examine what happens to a wave propagating in a finite material, while the material undergoes a slow index variation. In this scenario, only fields present inside the material during the entire transition are time-reversed. Thus, the finite size of the index-varying medium is what limits how slow the transition time of the refractive index can be, because the waves must be contained within the medium to be properly reversed. See simulations in [58].

Before closing, it is instructive to discuss the relation between the negative-frequency wave and phase conjugation, specifically in the context of the current results. In this context, thus far all known mechanisms for phase-conjugation could conjugate only the wave arriving at the phase-conjugating device (phase conjugate mirror, etc.), i.e., only the paraxial propagating modes are conjugated [53]. On the other hand, the perfect time-domain lens presented here gives rise to time-reversal for both propagating waves and evanescent waves comprising the spatial information (the image), which can have fine subwavelength structures. In principle, the perfect time-domain lens creates a time-reversed version of the EM fields with infinite spatial resolution. Another important difference is that phase conjugation typically gives rise only to a small frequency shift (via nondegenerate four wave mixing or similar mechanisms), implying that the slow-down in the time-reversed evolution is minute, whereas for the perfect time-lens the slow-down is arbitrarily large, limited only by the finite size of the medium.

To summarize, we presented the perfect time-domain lens, a device that perfectly reverses the temporal evolution of the EM fields. The perfect time-domain lens is created by incorporating negative index materials into the physics of time varying media. We showed that a time-boundary between a positive refractive index and a negative refractive index can create perfect time reversal of the temporal evolution of the EM fields, including the evanescent spectrum as well as the propagating spectrum. The reversed temporal evolution can be either accelerated or slowed down, allowing pulse compression, imaging and recording of ultrafast spatio-temporal information using a conventional slow detector. This ability sustains even when the index variation is arbitrarily slow, lifting the requirement for sub-cycle modulation for generation of negative-frequency waves. Finally, with recent experimental advances in the ability to significantly change in time the refractive index, both in microwaves [1,2] and at optical frequencies [3], the perfect time-domain lens can be demonstrated in the near future. The fact that the index modulation needs not be in the single cycle regime makes these phenomena much more accessible to experiments. An encouraging result is that a significant temporal change to the magnetic permeability was observed experimentally in GHz frequencies [54], making it promising for creating the first time-domain lens. Another candidate is using time-dependent non-Foster circuits [55,56], which were suggested [52,57] in other contexts. We focused here on EM waves, however, perfect time-domain lens can be created in any wave system where the wave impedance and propagation constant can be changed in time. This work suggests exploring "time-reversal" physics also in slowly-responding wave systems, which were believed not to respond to ultrafast phenomena.

This work was funded by the US Air Force Office for Scientific Research (AFOSR) and by the MAPATS program of the Israel Science Foundation.

References

[1]     H. Moussa, G. Xu, S. Yin, E. Galiffi, Y. Ra'di, and A. Alù, Observation of temporal reflection and broadband frequency translation at photonic time interfaces, Nat. Phys. **19**, 863 (2023).
[2]     T. R. Jones, A. V. Kildishev, M. Segev, and D. Peroulis, Time-reflection of microwaves by a fast optically-controlled time-boundary, Nat Commun **15**, 6786 (2024).
[3]     E. Lustig et al., Time-refraction optics with single cycle modulation, Nanophotonics **12**, 2221 (2023).
[4]     F. R. Morgenthaler, Velocity Modulation of Electromagnetic Waves, IRE Transactions on Microwave Theory and Techniques **6**, 167 (1958).
[5]     J. T. Mendonça and P. K. Shukla, Time Refraction and Time Reflection: Two Basic Concepts, Phys. Scr. **65**, 160 (2002).
[6]     F. Biancalana, A. Amann, A. V. Uskov, and E. P. O'Reilly, Dynamics of light propagation in spatiotemporal dielectric structures, Phys. Rev. E **75**, 046607 (2007).
[7]     H. J. Gerritsen, NONLINEAR EFFECTS IN IMAGE FORMATION, Applied Physics Letters **10**, 239 (1967).
[8]     D. L. Sounas and A. Alù, Non-reciprocal photonics based on time modulation, Nature Photon **11**, 774 (2017).
[9]     A. Akbarzadeh, N. Chamanara, and C. Caloz, Inverse prism based on temporal discontinuity and spatial dispersion, Opt. Lett., OL **43**, 3297 (2018).
[10]    V. Pacheco-Peña and N. Engheta, Temporal aiming, Light Sci Appl **9**, 129 (2020).
[11]    H. Li, S. Yin, E. Galiffi, and A. Alù, Temporal Parity-Time Symmetry for Extreme Energy Transformations, Phys. Rev. Lett. **127**, 153903 (2021).
[12]    O. Schiller, Y. Plotnik, O. Segal, M. Lyubarov, and M. Segev, Time-Domain Bound States in the Continuum, Phys. Rev. Lett. **133**, 263802 (2024).
[13]    J. R. Zurita-Sánchez, P. Halevi, and J. C. Cervantes-González, Reflection and transmission of a wave incident on a slab with a time-periodic dielectric function $\epsilon(t)$, Phys. Rev. A **79**, 053821 (2009).
[14]    J. R. Reyes-Ayona and P. Halevi, Observation of genuine wave vector (k or β) gap in a dynamic transmission line and temporal photonic crystals, Applied Physics Letters **107**, 074101 (2015).
[15]    E. Lustig, Y. Sharabi, and M. Segev, Topological aspects of photonic time crystals, Optica, OPTICA **5**, 1390 (2018).
[16]    N. Konforty, M.-I. Cohen, O. Segal, Y. Plotnik, V. M. Shalaev, and M. Segev, Second harmonic generation and nonlinear frequency conversion in photonic time-crystals, Light Sci Appl **14**, 152 (2025).
[17]    R. Tirole, S. Vezzoli, D. Saxena, S. Yang, T. V. Raziman, E. Galiffi, S. A. Maier, J. B. Pendry, and R. Sapienza, Second harmonic generation at a time-varying interface, Nat Commun **15**, 7752 (2024).
[18]    S. Saha, S. Gurung, B. T. Diroll, S. Chakraborty, O. Segal, M. Segev, V. M. Shalaev, A. V. Kildishev, A. Boltasseva, and R. D. Schaller, *Third Harmonic Enhancement Harnessing Photoexcitation Unveils New Nonlinearities in Zinc Oxide*, arXiv:2405.04891.
Page 14 of 17